\documentclass[11pt,a4paper]{article}
\usepackage{jcappub}
\usepackage{graphics}

\newcommand{\antares}{\mbox{ANTARES}}

\newcommand{\gammaraybursts}{gamma-ray bursts}

\newcommand{\allmneu}{muon-(anti)neutrino}
\newcommand{\mneus}{muon-neutrinos}
\newcommand{\mneu}{muon-neutrino}
\newcommand{\antimneus}{muon-antineutrinos}
\newcommand{\eneutwo}{$E_{\nu}^{-2}$}
\newcommand{\eneu}{$E_{\nu}$}
\newcommand{\etwo}{$E^{-2}$}
\newcommand{\ngrbs}{40}

\newcommand{\axislabel}[1]      {\textsl{#1}}
\newcommand{\captionfont}[1]    {\textsl{#1}}

\notoc
\begin{document}

\title{First search for neutrinos in correlation with \gammaraybursts\
with the \antares\ neutrino telescope}

\author[a]{S.~Adri\'an-Mart\'inez},
\author[b]{I. Al Samarai},
\author[c]{A. Albert},
\author[d]{M.~Andr\'e},
\author[e]{M. Anghinolfi},
\author[f]{G. Anton},
\author[g]{S. Anvar},
\author[a]{M. Ardid},
\author[h]{A.C. Assis Jesus},
\author[h,1]{T.~Astraatmadja,\note{Also at University of Leiden, the Netherlands}},
% ...

\affiliation[a]{\scriptsize{Institut d'Investigaci\'o per a la Gesti\'o Integrada de les Zones Costaneres (IGIC) - Universitat Polit\`ecnica de Val\`encia. C/  Paranimf 1 , 46730 Gandia, Spain.}}\vspace*{0.15cm}
\affiliation[b]{\scriptsize{CPPM, Aix-Marseille Universit\'e, CNRS/IN2P3, Marseille, France}}\vspace*{0.15cm}
\affiliation[c]{\scriptsize{GRPHE - Institut universitaire de technologie de Colmar, 34 rue du Grillenbreit BP 50568 - 68008 Colmar, France }}\vspace*{0.15cm}
\affiliation[d]{\scriptsize{Technical University of Catalonia, Laboratory of Applied Bioacoustics, Rambla Exposici\'o,08800 Vilanova i la Geltr\'u,Barcelona, Spain}}\vspace*{0.15cm}
\affiliation[e]{\scriptsize{INFN - Sezione di Genova, Via Dodecaneso 33, 16146 Genova, Italy}}\vspace*{0.15cm}
\affiliation[f]{\scriptsize{Friedrich-Alexander-Universit\"at Erlangen-N\"urnberg, Erlangen Centre for Astroparticle Physics, Erwin-Rommel-Str. 1, 91058 Erlangen, Germany}}\vspace*{0.15cm}
\affiliation[g]{\scriptsize{Direction des Sciences de la Mati\`ere - Institut de recherche sur les lois fondamentales de l'Univers - Service d'Electronique des D\'etecteurs et d'Informatique, CEA Saclay, 91191 Gif-sur-Yvette Cedex, France}}\vspace*{0.15cm}
\affiliation[h]{\scriptsize{Nikhef, Science Park,  Amsterdam, The Netherlands}}\vspace*{0.15cm}
\affiliation[i]{\scriptsize{APC - Laboratoire AstroParticule et Cosmologie, UMR 7164 (CNRS, Universit\'e Paris 7 Diderot, CEA, Observatoire de Paris) 10, rue Alice Domon et L\'eonie Duquet 75205 Paris Cedex 13,  France}}\vspace*{0.15cm}
\affiliation[j]{\scriptsize{LAM - Laboratoire d'Astrophysique de Marseille, P\^ole de l'\'Etoile Site de Ch\^ateau-Gombert, rue Fr\'ed\'eric Joliot-Curie 38,  13388 Marseille Cedex 13, France }}\vspace*{0.15cm}
\affiliation[k]{\scriptsize{INFN - Sezione di Bologna, Viale Berti-Pichat 6/2, 40127 Bologna, Italy}}\vspace*{0.15cm}
\affiliation[l]{\scriptsize{Dipartimento di Fisica dell'Universit\`a, Viale Berti Pichat 6/2, 40127 Bologna, Italy}}\vspace*{0.15cm}
\affiliation[m]{\scriptsize{INFN - Sezione di Pisa, Largo B. Pontecorvo 3, 56127 Pisa, Italy}}\vspace*{0.15cm}
\affiliation[n]{\scriptsize{IFIC - Instituto de F\'isica Corpuscular, Edificios Investigaci\'on de Paterna, CSIC - Universitat de Val\`encia, Apdo. de Correos 22085, 46071 Valencia, Spain}}\vspace*{0.15cm}
\affiliation[o]{\scriptsize{INFN -Sezione di Roma, P.le Aldo Moro 2, 00185 Roma, Italy}}\vspace*{0.15cm}
\affiliation[p]{\scriptsize{Dipartimento di Fisica dell'Universit\`a La Sapienza, P.le Aldo Moro 2, 00185 Roma, Italy}}\vspace*{0.15cm}
\affiliation[q]{\scriptsize{Clermont Universit\'e, Universit\'e Blaise Pascal, CNRS/IN2P3, Laboratoire de Physique Corpusculaire, BP 10448, 63000 Clermont-Ferrand, France}}\vspace*{0.15cm}
\affiliation[r]{\scriptsize{G\'eoazur - Universit\'e de Nice Sophia-Antipolis, CNRS/INSU, IRD, Observatoire de la C\^ote d'Azur and Universit\'e Pierre et Marie Curie, BP 48, 06235 Villefranche-sur-mer, France}}\vspace*{0.15cm}
\affiliation[s]{\scriptsize{INFN - Sezione di Bari, Via E. Orabona 4, 70126 Bari, Italy}}\vspace*{0.15cm}
\affiliation[t]{\scriptsize{INFN - Laboratori Nazionali del Sud (LNS), Via S. Sofia 62, 95123 Catania, Italy}}\vspace*{0.15cm}
\affiliation[u]{\scriptsize{COM - Centre d'Oc\'eanologie de Marseille, CNRS/INSU et Universit\'e de la M\'editerran\'ee, 163 Avenue de Luminy, Case 901, 13288 Marseille Cedex 9, France}}\vspace*{0.15cm}
\affiliation[v]{\scriptsize{Univ Paris-Sud , 91405 Orsay Cedex, France}}\vspace*{0.15cm}
\affiliation[w]{\scriptsize{Kernfysisch Versneller Instituut (KVI), University of Groningen, Zernikelaan 25, 9747 AA Groningen, The Netherlands}}\vspace*{0.15cm}
\affiliation[x]{\scriptsize{Direction des Sciences de la Mati\`ere - Institut de recherche sur les lois fondamentales de l'Univers - Service de Physique des Particules, CEA Saclay, 91191 Gif-sur-Yvette Cedex, France}}\vspace*{0.15cm}
\affiliation[y]{\scriptsize{Dipartimento di Fisica dell'Universit\`a, Largo B. Pontecorvo 3, 56127 Pisa, Italy}}\vspace*{0.15cm}
\affiliation[z]{\scriptsize{Royal Netherlands Institute for Sea Research (NIOZ), Landsdiep 4,1797 SZ 't Horntje (Texel), The Netherlands}}\vspace*{0.15cm}
\affiliation[aa]{\scriptsize{Dr. Remeis-Sternwarte and ECAP, Universit\"at Erlangen-N\"urnberg,  Sternwartstr. 7, 96049 Bamberg, Germany}}\vspace*{0.15cm}
\affiliation[ab]{\scriptsize{Universiteit Utrecht, Faculteit Betawetenschappen, Princetonplein 5, 3584 CC Utrecht, The Netherlands}}\vspace*{0.15cm}
\affiliation[ac]{\scriptsize{Universiteit van Amsterdam, Instituut voor Hoge-Energie Fysika, Science Park 105, 1098 XG Amsterdam, The Netherlands}}\vspace*{0.15cm}
\affiliation[ad]{\scriptsize{Moscow State University,Skobeltsyn Institute of Nuclear Physics,Leninskie gory, 119991 Moscow, Russia}}\vspace*{0.15cm}
\affiliation[ae]{\scriptsize{INFN - Sezione di Catania, Viale Andrea Doria 6, 95125 Catania, Italy}}\vspace*{0.15cm}
\affiliation[af]{\scriptsize{Dipartimento di Fisica ed Astronomia dell'Universit\`a, Viale Andrea Doria 6, 95125 Catania, Italy}}\vspace*{0.15cm}
\affiliation[ag]{\scriptsize{D\'epartement de Physique Nucl\'eaire et Corpusculaire, Universit\'e de Gen\`eve, 1211, Geneva, Switzerland}}\vspace*{0.15cm}
\affiliation[ah]{\scriptsize{Institute for Space Sciences, R-77125 Bucharest, M\u{a}gurele, Romania     }}\vspace*{0.15cm}
\affiliation[ai]{\scriptsize{IPHC-Institut Pluridisciplinaire Hubert Curien - Universit\'e de Strasbourg et CNRS/IN2P3  23 rue du Loess, BP 28,  67037 Strasbourg Cedex 2, France}}\vspace*{0.15cm}
\affiliation[aj]{\scriptsize{ITEP - Institute for Theoretical and Experimental Physics, B. Cheremushkinskaya 25, 117218 Moscow, Russia}}\vspace*{0.15cm}
\affiliation[ak]{\scriptsize{Dipartimento di Fisica dell'Universit\`a, Via Dodecaneso 33, 16146 Genova, Italy}}\vspace*{0.15cm}
\affiliation[al]{\scriptsize{University Mohammed I, Laboratory of Physics
of Matter and Radiations, B.P.717, Oujda 6000, Morocco}}\vspace*{0.15cm}

\abstract{A search for neutrino-induced muons in correlation with a selection of
\ngrbs\ \gammaraybursts\ that occurred in 2007 has been performed with the
\antares\ neutrino telescope. 
During that period, the detector consisted of 5 detection lines.
The \antares\ neutrino telescope is sensitive to TeV--PeV neutrinos
that are predicted from \gammaraybursts.
No events were found in correlation with the prompt photon emission of the 
\gammaraybursts\ and upper limits 
have been placed on the flux and fluence of neutrinos for different models.}

\maketitle
\flushbottom

\section{Introduction}
\label{sec:introduction}
Gamma-ray bursts (GRBs), transient flashes of gamma-rays having a 
duration of sub-seconds up to several hundred seconds, are the most 
powerful known extra-galactic events. 
Several models predict a burst of high-energy neutrinos in concurrence 
with the flash of gamma-rays, also referred to as the prompt emission 
(see e.g. [1] for a review). 
In this, neutrinos are produced by interactions of protons that are 
accelerated by shock waves with energetic ambient photons.
A variety of models has been put forth covering a wide range in the expected
amount of neutrinos. 
Some of those models have already been challenged by
the recent IceCube measurements~\cite{ic59}. 
In particular, the IceCube results do not support the notion of 
GRBs as primary source of ultra high-energy cosmic rays, 
at least under the assumption that these ultra high-energy cosmic rays 
are protons coming from the decay of photohadronically produced neutrons 
(i.e. via the resonant channel 
$p+\gamma\rightarrow\Delta^{+}\rightarrow n+\pi^{+}$) and are 
therefore connected to neutrinos. 
In this scenario, neutrons, as opposed to the protons, 
can escape because they are not magnetically confined inside the source.
Nevertheless, a viable phase space of models with a potentially 
measurable neutrino flux from GRBs in the TeV--PeV region can still 
be tested for a better understanding of the nature of these
energetic events.

In this paper, the first data taken with the \antares\ neutrino
telescope in 2007 are used to perform a search for TeV--PeV neutrinos in
correlation with a selection of GRBs detected by satellite observatories.
The data are treated in a stacking approach in which the data observed
during the prompt emission of all the selected GRBs are accumulated.
The detection of a single neutrino event would constitute an
observation with more than 3$\sigma$ significance. 

Previous searches for neutrinos from GRBs have been performed in the Northern 
hemisphere by AMANDA~\cite{amanda} and IceCube~\cite{ic22,ic40} at similar energies 
(TeV--PeV) and by ANITA~\cite{anita} at higher energies ($>$PeV). 
IceCube also included some GRBs in the Southern hemisphere~\cite{ic59} that is 
observed by \antares\ but not during the period considered in this analysis.
Other searches in the Southern hemisphere were performed by 
Super-Kamiokande~\cite{sk} at lower energies (MeV--100 TeV) 
and by the Baikal neutrino telescope NT200~\cite{baikal}.
Both these experiments have a lower sensitivity than the current analysis in the 
comparable energy range. 
The \antares\ neutrino telescope is the most sensitive instrument to observe 
high-energy neutrinos from the GRBs considered in this analysis.
In addition, the vast majority of these GRBs have not been studied for neutrino emission 
in the TeV--PeV range before.
Despite the fact that more stringent limits have been published~\cite{ic59,ic22,ic40} 
for different GRB samples, \antares\ might still have observed events
from the GRB sample presented here.

The completed \antares\ detector is the largest neutrino telescope on the 
Northern hemisphere and is sensitive to neutrinos in the TeV to PeV energy range.
Located in the Mediterranean Sea, the \antares\ detector is sensitive 
to GRBs in the Southern hemisphere where the sensitivity of IceCube
significantly suffers from the large background from muons produced by
cosmic ray interactions in the atmosphere above the detector.
Due to the transient nature of the GRBs and the variety of their
characteristics, it is essential to permanently monitor the full sky
in order to maximise the probability to observe a neutrino signal. 
The search described in this paper is the first in a
series of searches which will be pursued with the data that will be 
acquired by the completed \antares\ detector up to 2016
and then continued with data from the future KM3NeT~\cite{km3net}
neutrino observatory in the Mediterranean Sea.

\section{Neutrino detection}
\label{sec:neu_detection}
High-energy neutrinos can be detected indirectly by a neutrino
telescope such as \antares.
The detection principle relies on measuring the Cherenkov light
induced by high-energy charged particles that are produced in a
neutrino interaction inside or near the instrumented volume.
In particular, a high-energy muon is produced in a charged current
muon-neutrino interaction.
High-energy muons can travel large distances, which facilitates an
accurate determination of the direction.
At the typical neutrino energies considered in this
analysis (5~TeV--6~PeV), the direction of the muon closely follows that of the
incident neutrino.
Above 5~TeV the median angle between the neutrino direction and the 
muon direction is less than $0.3^{\circ}$.
The direction of a high-energy muon can thus be correlated to the positions 
of GRBs. 

\begin{figure}[t]
\begin{picture}(10,10.5)
\put(0.0,0.0){\scalebox{1.0}{
        \put(5,0){\scalebox{0.4}{\includegraphics{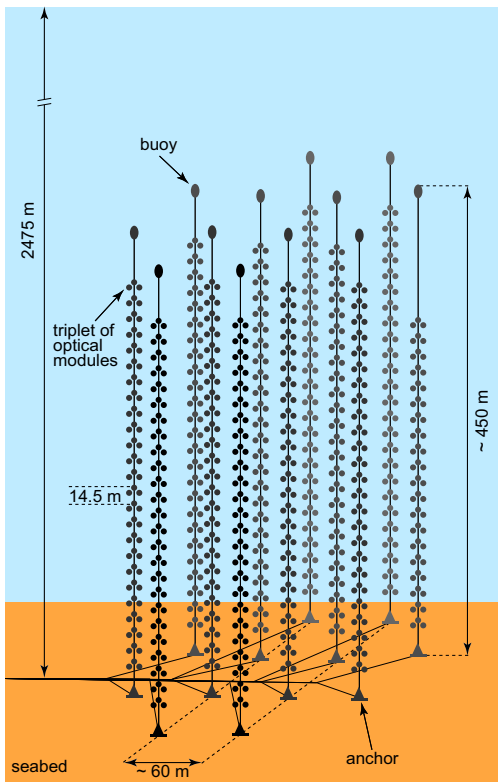}}}
}}
\end{picture}
\caption{\captionfont{Schematic view of the \antares\ detector in its final configuration.}} 
\label{fig:detector}
\end{figure}

The Cherenkov light is detected by photo-multiplier tubes, housed in
optical modules~\cite{om}.  
Triplets of optical modules are attached to vertical detection lines of
about 450 metres height, which are anchored to the sea bed at a depth of 2475
metres and held upright by a buoy. 
In its final configuration, the \antares\ detector consists of 12
detection lines with a spacing of about 60 metres (see figure~\ref{fig:detector}).
Eleven of these detection lines each contain 25 evenly spaced triplets
of optical modules and one contains 20 triplets.
The positions and orientations of the optical modules vary due to
the sea currents.
An acoustic positioning system, combined with
compasses and tiltmeters located along the detection lines, measures
the positions and orientations of the optical modules with an
accuracy of about 10~cm.
A detailed description of the detector is given in
ref.~\cite{detector}. 

The arrival time and charge of the photo-multiplier tube signals are
digitised~\cite{ars} and transmitted to shore.
The absolute time stamping is performed by interfacing the clock
system to the GPS which provides a time accuracy of about 100~ns with
respect to Universal Time Coordinated~\cite{timecalibration}.
The data are dominated by optical background due to bioluminescence
and natural radioactive decays.
On shore, the physics signals are selected from the data stream by a
software data filter that operates in real time.
This data filter has multiple algorithms implemented, each designed to
find a specific physics signal.
The physics events are stored on disk for offline reconstruction.
A detailed description of the data acquisition and data filtering
is given in ref.~\cite{daq}.

\section{Data selection}
\label{sec:data_selection}
The analysis described in this paper has been applied to data
collected in 2007, in the period from January 27 to December 7.
At this time, the \antares\ detector was still under construction and
consisted of 5 detection lines.

\subsection{GRB selection}
\label{subsec:grb_selection}
The GRBs examined for neutrino emission were selected from
the observations of GRBs by satellite instruments, as archived by the 
Gamma-ray Burst Coordinates Network
(GCN)~\cite{gcncircular,gcnreport}.
In order to suppress the background from muons produced by
cosmic ray interactions in the atmosphere above the detector,
hereafter referred to as atmospheric muons,
GRBs were selected that occurred below the horizon of the \antares\
detector. 
Neutrinos from such GRBs
would traverse the Earth and cross the detector in an upward-going direction.
In the period considered, 46~GRBs occurred below the \antares\ horizon
during physics data taking.

By requiring the availability of detector alignment data, and by
applying quality criteria based on the 
environmental conditions, the data in coincidence with 6 GRBs were
excluded.
The time and position information of the \ngrbs\ remaining GRBs, listed in
table~\ref{tab:grbs}, were used to search for a correlated neutrino
signal.
A total of 32~GRBs were detected by the Swift
satellite~\cite{swift}, 4 by INTEGRAL~\cite{integral} and 4 by other
satellites of the Third Interplanetary Network~\cite{ipn}.
\begin{table}[t]
\begin{tabular}{lll|} \hline
GRB       & GCN Circular & GCN Report \\ \hline
GRB070207 & 6089         &       \\ 
GRB070209 & 6101         & 32.2  \\ 
GRB070227 &              & 37.1  \\ 
GRB070311 & 6189         &       \\ 
GRB070326 & 6653         &       \\ 
GRB070328 &              & 42.3  \\ 
GRB070330 &              & 43.2  \\ 
GRB070419 &              & 47.1  \\ 
GRB070419B&              & 49.1  \\ 
GRB070420 &              & 48.1  \\ 
GRB070429B& 7140         & 51.1  \\ 
GRB070508 &              & 54.2  \\ 
GRB070509 &              & 55.1  \\ 
GRB070517 &              & 56.2  \\ 
GRB070611 &              & 63.3  \\ 
GRB070612 & 6556         & 64.1  \\ 
GRB070612B&              & 65.1  \\ 
GRB070615 & 6537         &       \\ 
GRB070707 & 6615         &       \\ 
GRB070721 &              & 72.2  \\ \hline
\end{tabular}
\begin{tabular}{lll} \hline
GRB       & GCN Circular & GCN Report \\ \hline
GRB070724 &              & 74.2  \\ 
GRB070729 &              & 77.1  \\ 
GRB070805 & 6711         &       \\ 
GRB070808 & 6724         &       \\ 
GRB070810B&              & 81.1  \\ 
GRB070821 & 6766         &       \\ 
GRB070824 & 6768         &       \\ 
GRB070911 & 6810         & 82.2  \\ 
GRB070913 &              & 88.1  \\ 
GRB070923 & 6821         &       \\ 
GRB071001 &              & 86.1  \\ 
GRB071003 &              & 87.2  \\ 
GRB071008 &              & 91.1  \\ 
GRB071010 & 6864         & 89.2  \\ 
GRB071010B& 6888         & 92.1  \\ 
GRB071017 & 6927         &       \\ 
GRB071028B&              & 105.1 \\ 
GRB071031 &              & 99.1  \\ 
GRB071112B&              & 103.1 \\ 
GRB071117 &              & 106.2 \\ \hline
\end{tabular}
\caption{\captionfont{List of the GRBs that are included in the
analysis. The column ``GRB" indicates the name of the GRB. The columns
``GCN Circular" and ``GCN Report" indicate the numbers of the GCN
Circulars archive~\cite{gcncircular} and GCN Reports~\cite{gcnreport}
respectively, from which the measured GRB data were taken.}}
\label{tab:grbs}
\end{table}

\subsection{Data processing}
\label{subsec:data_processing}
This analysis focuses on the search for (anti)muons, produced by
\allmneu\ charged current interactions.
Throughout the rest of this paper, \mneus\ denotes both \mneus\ and \antimneus.
A data filter algorithm that is sensitive to muons from any direction
has been applied during data taking.
This algorithm first selects photons detected within 20~ns by separate optical 
modules in the same triplet.
It then identifies a muon by requiring that for at least 5 triplets 
the relative arrival times of these photons
are compatible with the signal expected from a muon traversing the detector.
The muon purity for this data filter algorithm is better than 90\%.
Its average event rate was 1.0~Hz in the period considered
and is mainly due to the background from atmospheric muons. 

These events were reconstructed offline to determine the muon
trajectory, using a multi-stage fitting procedure.
The reconstruction code follows the algorithm described in ref.~\cite{aart_proefschrift}. 
Minor modifications were made to improve agreement between data and Monte
Carlo. 
The most important modification compared to ref.~\cite{aart_proefschrift} is that the amplitude
information of the detected photons (hereafter referred to as hits) is discarded. 
As a result, the initial selection of signal hits
is purely based on coincidences and the causality criterion.
The algorithm starts with a linear prefit which
is used as a starting track for the subsequent stages. 
In addition, eight different starting tracks are generated by rotating and translating 
the result of the prefit.
This is to increase the probability to find the global maximum of the
likelihood function.
The final stage of the fitting procedure consists of a maximum likelihood fit
of the measured photon arrival times.
A quality parameter for the fit, indicated by $\Lambda$, is determined based on
the final value of the likelihood function.
The $\Lambda$ parameter is quantified by 
\begin{equation}
 \Lambda \equiv \frac{\log(L)}{ N_{\rm hits}-5} + 0.1 \times (N_{\rm comp} -1 ),
\label{eq:lambda}
\end{equation}
which incorporates the maximum value of the likelihood, $L$, and the number of degrees of
freedom of the fit, i.e. the number of hits, $N_{hits}$, used in the fit minus the number of fit
parameters;
$N_{comp}$ is the number of times the repeated initial steps of the reconstruction
converged to the same result. 
In general, $N_{comp}$~=~1 for badly reconstructed events while
it can be as large as nine for well reconstructed events. 
The coefficient 0.1 in eq.~\ref{eq:lambda} was chosen to maximise the separation
in $\Lambda$ between simulated signal and misreconstructed
downgoing muons.
The same reconstruction algorithm was used for the analysis described in 
ref.~\cite{aart_pointsource2} which contains a brief description of the algorithm.
For a more detailed description of the reconstruction algorithm see 
ref.~\cite{aart_proefschrift}.

\begin{figure}[t]
\begin{picture}(10,7)
\put(0.0,0.0){\scalebox{1.0}{
        \put(0.5,0.4){\scalebox{0.8}{\includegraphics{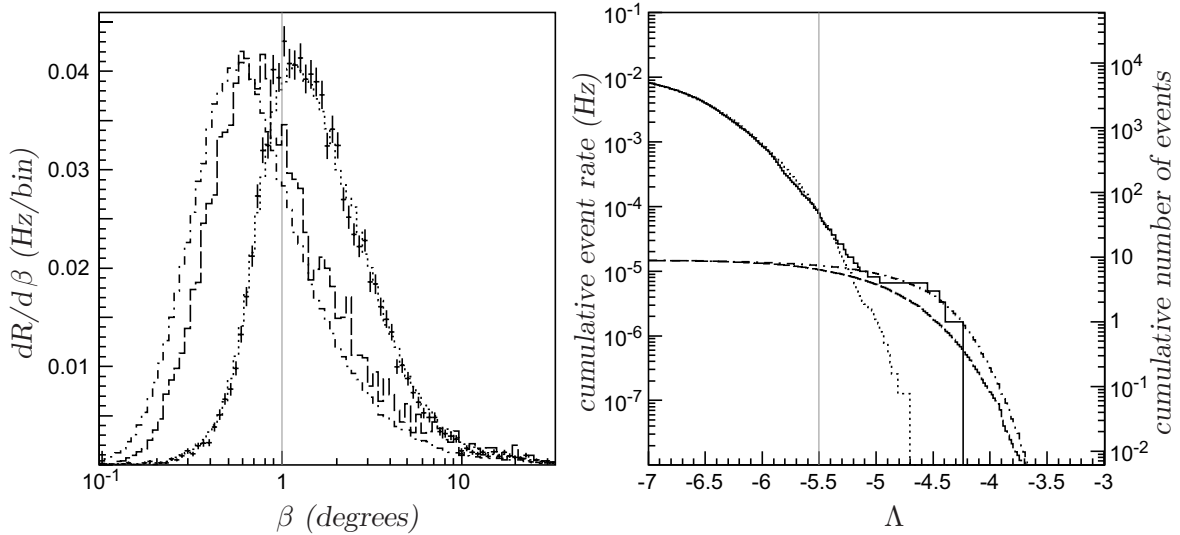}}}
        \put(0.0,2.2){\rotatebox{90}{\axislabel{dR/d\,$\beta$ (Hz/bin)}}}
        \put(7.4,1.5){\rotatebox{90}{\axislabel{cumulative event rate (Hz)}}}
        \put(15.0,1.2){\rotatebox{90}{\axislabel{cumulative number of events}}}
        \put(3.5,0){\axislabel{$\beta$ (degrees)}}
        \put(11.5,0){\axislabel{$\Lambda$}}
}}
\end{picture}
\caption{\captionfont{Left: Distribution of the estimated angular uncertainty
$\beta$ on the direction of the reconstructed muon track for upward
reconstructed simulated atmospheric muons (dotted line), atmospheric
\mneus$\,\times\,1300$ (dashed line) and signal events with an \etwo\
spectrum (dash-dotted line) normalised to the atmospheric \mneus,
compared to upward reconstructed 
events in the data covering about 7 days and containing the \ngrbs\
selected GRBs (data points), without further cuts. Right:
Cumulative $\Lambda$ distribution for upward reconstructed simulated
atmospheric muons (dotted line), atmospheric \mneus\ (dashed
line) and signal events with an \etwo\
spectrum (dash-dotted line) normalised to the atmospheric \mneus,
 compared to upward reconstructed events in the data covering
about 7 days and containing the \ngrbs\ selected GRBs (solid line),
for events with $\beta\leq1^{\circ}$. The vertical lines indicate the
analysis cuts.}} 
\label{fig:aafit}
\end{figure}

The fit also provides an estimate of the uncertainties on the track parameters.
These are used to select events with a well defined direction.
The distribution of the estimated angular uncertainty on the direction of the reconstructed 
muon track, indicated by $\beta$, is shown in figure~\ref{fig:aafit}
(left).
Since all selected GRBs occurred below the \antares\ horizon, only
events with an upward reconstructed direction are considered.

Due to the changing detector conditions during the period considered, and the 
varying environmental conditions, the statistics of the uncorrelated data
that are equivalent to the conditions during the GRBs is too limited to estimate the background.
Hence the background estimate has been made based on simulations.
Figure~\ref{fig:aafit} includes the expected distributions from
atmospheric muons, atmospheric \mneus\ as well as an assumed signal
with an \etwo\ spectrum, obtained from simulations.
The neutrino signal is generated with software packages~\cite{david}
that simulate the neutrino interaction as well as the production and
propagation of charged particles.
The simulation uses the model for the atmospheric \mneu\ flux
from ref.~\cite{bartol}.
The atmospheric muon contribution is simulated with the MUPAGE
package~\cite{mupage}, which is based on a full Monte Carlo simulation
of primary cosmic ray interactions and shower propagation in the 
atmosphere and reproduces the MACRO data~\cite{macro1,macro2}.
The simulated atmospheric muon contribution has an equivalent 
live time of one month.

In the simulations, the stochastic energy loss of the muons, the
production and propagation of the Cherenkov photons, the response of
the photo-multiplier tubes to Cherenkov light and the simulation of
the detector electronics are all included. 
The simulated photon signals are processed with the same data filter
and reconstruction algorithms as the data.
To obtain a realistic simulation of the varying environmental
conditions due to bioluminescence, the measured optical background in
coincidence with the prompt emission of each of the selected GRBs is
taken from the data and added to the simulated events.

Selection cuts on the $\Lambda$ and $\beta$ parameters
are set to achieve a reliable rejection of misreconstructed
atmospheric muons, while keeping a high signal efficiency.
Neutrino candidates are required to be reconstructed as upgoing muons,
with an estimated angular uncertainty $\beta\leq1^{\circ}$, see
figure~\ref{fig:aafit} (left). 
This cut removes 72\% of the misreconstructed atmospheric muons.
Figure~\ref{fig:aafit} (right) shows the cumulative
$\Lambda$ distribution for upgoing reconstructed events, where the cut
of $\beta\leq1^{\circ}$ has been applied.
Neutrino candidates are, in addition, required
to have a quality value of $\Lambda\geq-5.5$.
The cuts on the $\Lambda$ and $\beta$ parameters are chosen such that
the background rate is reduced to a level below
$2\times10^{-5}$~Hz\,sr$^{-1}$, while the signal efficiency is about
60\% for the models considered in this analysis.
The strong constraints from the time and direction coincidence with a
GRB allow for a looser cut on the quality parameter $\Lambda$ than
that applied for the \antares\ point source search~\cite{aart_pointsource}.

In both plots of figure~\ref{fig:aafit}, the contribution from
atmospheric muons was scaled by 0.85 to reproduce the data.
This scaling factor is well within the uncertainty
on the flux normalisation~\cite{mupage_line5}.

\begin{figure}[t]
\begin{picture}(10,7.5)
\put(0.0,0){\scalebox{1.0}{
        \put(0.5,0.3){\scalebox{0.65}{\includegraphics{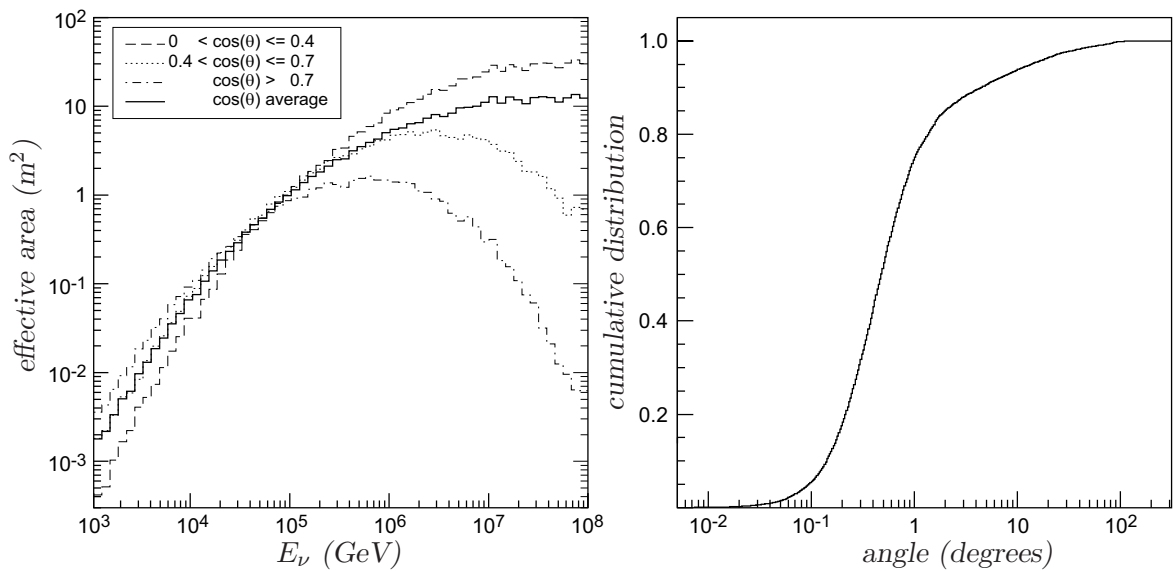}}}
        \put(0.0,2.5){\rotatebox{90}{\axislabel{effective area (m$^2$)}}}
        \put(7.8,2.0){\rotatebox{90}{\axislabel{cumulative distribution}}}
        \put(3.5,0){\axislabel{$E_{\nu}$ (GeV)}}
        \put(11.2,0){\axislabel{angle (degrees)}}
}}
\end{picture}
\caption{\captionfont{Left: the angle averaged neutrino effective area
of the detector consisting of 5 detection lines, and the effective
areas for different neutrino zenith angle bins
(cos($\theta$)\,=\,1 corresponds to vertically upward).
The selection cuts are included.
Right: the cumulative distribution of the angle between the reconstructed
muon direction and the true neutrino direction for the detector consisting of
5 detection lines. 
The distribution is shown for selected events with an \eneutwo\ spectrum.}}
\label{fig:effective_area}
\end{figure}

The neutrino effective area of the detector consisting of 5 detection 
lines is shown in figure~\ref{fig:effective_area} (left)
as a function of the neutrino energy.
The presented effective areas include the selection cuts and are the 
average of the effective areas for \mneus\ and \antimneus.
The angle averaged effective area is shown as well as the 
effective areas for different neutrino zenith angle bins.
For vertically upward-going neutrinos the effective area is suppressed 
at high energies due to the absorption in the Earth.

The angular resolution of the detector is determined by the angular uncertainty 
on the reconstructed muon direction ($\beta$) and the neutrino scattering angle. 
The distribution of the angle between the reconstructed muon direction
and the true neutrino direction for the detector consisting of 5 detection 
lines was evaluated with simulations.
The cumulative distribution of this angle is shown in 
figure~\ref{fig:effective_area} (right) for events with a neutrino 
spectrum proportional to \eneutwo\ that passed the selection cuts.
The median of this angular resolution was estimated to be $0.5\pm0.1$
degrees.

\section{Data analysis}
\label{sec:detector_performance}
\begin{figure}[t]
\begin{picture}(10,10.5)
\put(0.0,0){\scalebox{1.0}{
        \put(0.55,0.4){\scalebox{0.8}{\includegraphics{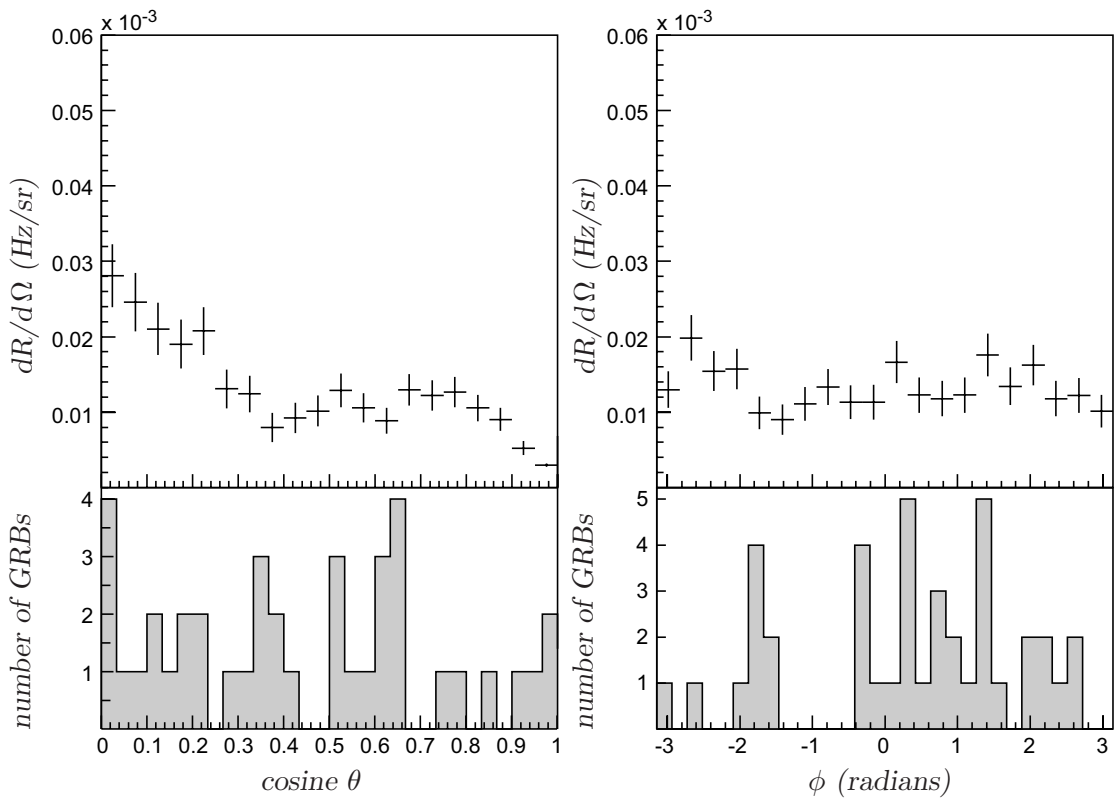}}}
        \put(0.0,5.5){\rotatebox{90}{\axislabel{dR/d\,$\Omega$ (Hz/sr)}}}
        \put(7.4,5.5){\rotatebox{90}{\axislabel{dR/d\,$\Omega$ (Hz/sr)}}}
        \put(10.5,0){\axislabel{$\phi$ (radians)}}
        \put(3.3,0){\axislabel{cosine $\theta$}}
        \put(0.0,0.8){\rotatebox{90}{\axislabel{number of GRBs}}}
        \put(7.4,0.8){\rotatebox{90}{\axislabel{number of GRBs}}}
}}
\end{picture}
\caption{\captionfont{Top: the estimated background rate per
unit solid angle as a function of the reconstructed muon direction in
detector coordinates (left: zenith angle cosine $\theta$, right:
azimuthal angle $\phi$ in radians; cos($\theta$)\,=\,1 corresponds to
vertically upward). Bottom: the location of each of
the \ngrbs\ selected GRBs.}}
\label{fig:bkg}
\end{figure}

The remaining background due to atmospheric \mneus\ (89\%) and 
misreconstructed atmospheric muons (11\%), after the cuts on $\beta$ and
$\Lambda$, was estimated from simulations.
The measured optical background from the data 
in coincidence with the prompt emission of each of the selected GRBs
was added to the simulated events.
The background rate per unit solid angle as a function of the
reconstructed muon direction in detector coordinates is shown in
figure~\ref{fig:bkg} (top plots).
The contribution from atmospheric muons was scaled by 0.85 to
reproduce the data.
The background rate as a function of the reconstructed zenith angle,
$\theta$, is less isotropic compared to the background rate as a
function of the reconstructed azimuthal angle, $\phi$.
However, since the 
selected GRBs are distributed rather isotropically 
(bottom plots of figure~\ref{fig:bkg}), the average background rate
may be used for each GRB.
The expected background rate per unit solid angle is estimated by
taking a weighted average of the background rate in each solid angle
bin, where the weighting accounts for the relative duration of GRBs in
that solid angle bin.
This results in an estimated background rate of
$1.54\times10^{-5}$~Hz\,sr$^{-1}$.

A neutrino candidate is considered to come from the
GRB when the 
reconstructed muon track points back to the 
GRB within $2^{\circ}$.
From simulations it was estimated that for about 85\% of the neutrino
candidates, the muon is reconstructed within $2^{\circ}$ from the
neutrino direction. 
This high signal efficiency is the result of the selection of
events with a good angular resolution (the cut on the $\beta$
parameter).
The uncertainty on the source positions of the GRBs considered
(table~\ref{tab:grbs}) is smaller than $0.07^{\circ}$.
The total background, $\mu_{bkg}$, due to atmospheric \mneus\ and
misreconstructed atmospheric muons during the prompt emission of the
\ngrbs\ GRBs can be expressed as
\begin{equation}
\mu_{bkg}~=~\frac{dR}{d\Omega}\times\Omega\times T
\end{equation}
where $\frac{dR}{d\Omega}$ is the background rate per unit solid
angle, $\Omega$ is the solid angle of the 2$^\circ$ search cone and $T$
is the total prompt emission duration of the \ngrbs\ GRBs.
For the prompt emission duration of the GRB, the so called $T_{90}$ time
interval is taken~\cite{kouveliotou}, extended by 5\% before the start,
and 5\% after the end of this time interval. 
$T$ amounts to 2114~seconds.
The total estimated background amounts to $1.24\times10^{-4}$ events.

\section{Results}
\label{sec:results}
This analysis focuses only on the detection of \mneus\ in
coincidence with the prompt emission phase of the GRB.
The search for neutrinos in correlation with GRBs was done in a
stacking analysis, in which all data in coincidence with the prompt
emission of the GRBs were accumulated.

Selected neutrino candidates are considered to be correlated with
a GRB when their detection time is in coincidence with the prompt
emission of the GRB, assuming that neutrinos travel at the speed of light.
After unblinding the data, no neutrino candidates were found in
correlation with the selected GRBs.

\subsection{GRB neutrino spectra}
\label{subsec:neu_spectra}
A general neutrino spectrum of \eneutwo\ is assumed for the neutrino
emission from GRBs, where \eneu\ is the neutrino energy.
In addition, three other energy
spectra for neutrino emission from GRBs have been considered:
the energy spectrum proposed by Guetta et al.\ \cite{guetta3},
the general Waxman and Bahcall energy spectrum~\cite{wb2}, and the
energy spectrum proposed by Ahlers et al.\ \cite{ahlers}.

\begin{figure}[t]
\begin{picture}(10,7.5)
\put(0.0,0.0){\scalebox{1.0}{
        \put(0.5,0.5){\scalebox{0.8}{\includegraphics{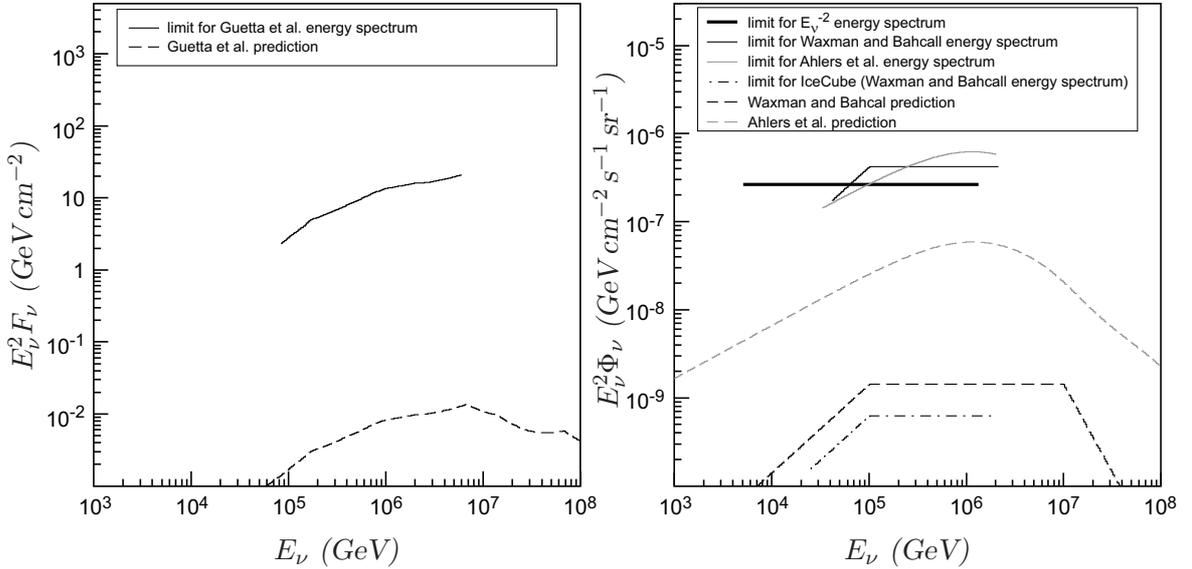}}}
        \put(0.0,2.5){\rotatebox{90}{\axislabel{$E_{\nu}^{2}F_{\nu}$ (GeV\,cm$^{-2}$)}}}
        \put(7.7,2.0){\rotatebox{90}{\axislabel{$E_{\nu}^{2}\Phi_{\nu}$ (GeV\,cm$^{-2}$\,s$^{-1}$\,sr$^{-1}$)}}}
        \put(3.5,0.0){\axislabel{$E_{\nu}$ (GeV)}}
        \put(11.0,0.0){\axislabel{$E_{\nu}$ (GeV)}}
}}
\end{picture}
\caption{\captionfont{Left: the 90\% CL upper limit on the \mneu\
fluence $F_{\nu}$ from the \ngrbs\ selected GRBs for the energy 
spectrum according to Guetta et al.\ (solid line) and the  
corresponding total estimated \mneu\ fluence including
oscillations (dashed line).
Right: the 90\% CL upper limit on the diffuse \mneu\ flux $\Phi_{\nu}$
for the \eneutwo\ energy spectrum (thick solid line), the Waxman and Bahcall 
energy spectrum (thin black line), and the energy spectrum according to 
Ahlers et al.\ (thin grey line).
The black dash-dotted line is the 90\% CL upper limit of IceCube~\cite{ic59} 
assuming the Waxman and Bahcall energy spectrum.
Also shown are the estimated diffuse \mneu\ fluxes including oscillations
assuming the Waxman and Bahcall energy spectrum (black dashed line) and 
that according to Ahlers et al.\ (grey dashed line).}}
\label{fig:limits}
\end{figure}

The energy spectrum according to Guetta et al.\ is calculated for each
GRB individually using the data from the instruments on the
satellite that detected the GRB, taken from the references given in
table~\ref{tab:grbs}.
The method to calculate a neutrino fluence from an individual GRB is
provided in ref.~\cite{guetta3}.
The prediction therein referred to as "Model 2" is used.
In case not all required parameters were measured, nor default values
are provided in ref.~\cite{guetta3}, the values as listed in
table~\ref{tab:guetta_defaults} were used.
\begin{table}[t]
\begin{tabular}{ll} \hline
Parameter           & Value                           \\ \hline
$z$                 & 0.25 or 2.8                      \\ 
$\epsilon_{e}$      & 0.1                              \\ 
$\epsilon_{B}$      & 0.1                              \\ 
$\varepsilon_{\gamma}^{b}$ & 0.4 MeV                          \\ 
$\alpha_{\gamma}$          & 1                                \\ 
$\beta_{\gamma}$           & 2                                \\ 
$L_{\gamma}$               & $10^{51}$~ergs\,s$^{-1}$          \\ 
$F_{\gamma}$               & $6\times10^{-6}$~ergs\,cm$^{-2}$ \\ \hline
\end{tabular}
\caption{\captionfont{Assumed parameter values used for
calculating the neutrino spectrum for an individual GRB using the
method described in ref.~\cite{guetta3} (see text). 
The symbols correspond to those used in ref.~\cite{guetta3}.
The redshift $z$ is set to 0.25 for GRBs with a prompt
emission duration of less than 2~s~\cite{redshift_short}, and set to 2.8 for
other GRBs~\cite{redshift_long}; 
the values for the fraction of the internal energy in
electrons ($\epsilon_{e}$) and the magnetic field ($\epsilon_{B}$) are
taken from~\cite{waxman}; the default values for the break energy in the
$\gamma$-ray spectrum $\varepsilon_{\gamma}^{b}$, the spectral indices of
the $\gamma$-ray spectrum before ($\alpha_{\gamma}$) and after
($\beta_{\gamma}$) $\varepsilon_{\gamma}^{b}$, the $\gamma$-ray
luminosity $L_{\gamma}$, and the $\gamma$-ray fluence $F_{\gamma}$ are
the mean values from the Swift catalogue~\cite{swift1,swift2}.}}
\label{tab:guetta_defaults}
\end{table}
Within the list of GRBs no single GRB yields a detectable signal.
The total estimated \mneu\ fluence for the \ngrbs\ selected GRBs
according to Guetta et al.\ is the sum of the calculated individual
\mneu\ fluences, and is shown in figure~\ref{fig:limits} (left).
The total number of expected events from the \ngrbs\ selected GRBs for
the estimated \mneu\ fluence with the energy spectrum according to
Guetta et al.\ is $1.7\times10^{-3}$.
The estimated \mneu\ fluence represents the
fluence at Earth and includes the effect of neutrino oscillations.

Waxman and Bahcall~\cite{wb2} provide an approximate estimate of the
neutrino energy spectrum, which is assumed to be the same for each
GRB.
The expected number of events for \ngrbs\ GRBs assuming the Waxmann 
and Bahcall energy spectrum is $7.0\times10^{-3}$.
The estimated GRB \mneu\ intensity assuming a Waxman and Bahcall 
energy spectrum is presented as a diffuse flux in figure~\ref{fig:limits}
(right) where a total of $10^3$ GRBs are expected per year.\footnote{In
ref.~\cite{wb2} the evolution correction is assumed to be 1.}
The right plot in figure~\ref{fig:limits} also includes an estimated
diffuse \mneu\ flux according to Ahlers et al.\ \cite{ahlers} where
typical values for the parameters of the GRB environment were used as 
provided in ref.~\cite{ahlers}.
The estimated diffuse \mneu\ fluxes for both energy spectra 
represent the fluxes at Earth and include the effect of 
neutrino oscillations.

\subsection{Upper limits on the diffuse neutrino flux and the neutrino
fluence}
\label{subsec:upper_limits}
Since no events were found in correlation with the prompt photon
emission of the \ngrbs\ selected GRBs, upper limits have been placed
on the intensity of the diffuse \mneu\ flux and the \mneu\ fluence
at Earth for the different models.
The 90\% confidence level limits were set using the Feldman-Cousins
prescription~\cite{feldman_cousins}, and are shown in
figure~\ref{fig:limits}.
The same systematic uncertainties as
described in ref.~\cite{aart_pointsource} have been considered.
These include the effect of reduced optical
module efficiencies and the effects which have a net result of
degrading the time resolution, such as possible mis-alignments of the
detector, inaccuracies in the simulation of light propagation in the
water or in the transit time distribution of the photo-multiplier tubes.
The impact of the systematic uncertainties were evaluated by including these
effects in the simulation, described in ref.~\cite{aart_pointsource}.
This results in a degradation of the limits of less than 10\%. 

For the \eneutwo\ energy spectrum, 90\% of the signal is expected in
the energy range 5.2~TeV\,$<\,E_{\nu}\,<$\,1.4~PeV.
The upper limit on the diffuse \mneu\ flux for the 
\eneutwo\ energy spectrum is
$2.7\times10^{-7}$ ($E_{\nu}$/GeV)$^{-2}$\,GeV$^{-1}$\,cm$^{-2}$\,s$^{-1}$\,sr$^{-1}$.
For the energy spectrum according to Waxman and Bahcall 90\% of the
signal is expected in the energy range
41~TeV\,$<\,E_{\nu}\,<$\,2.1~PeV
and a 90\% CL upper limit of about 294 times the predicted flux as
shown in figure~\ref{fig:limits} (right) was set.
For the energy spectrum according to Ahlers et al.\ 90\% of the
signal is expected in the energy range
33~TeV\,$<\,E_{\nu}\,<$\,2.0~PeV and a 90\% CL upper limit of about 11
times the predicted flux as shown in figure~\ref{fig:limits} (right) was set.
For the energy spectrum according to Guetta et al.\ 90\% of the
signal is expected in the energy range
86~TeV\,$<\,E_{\nu}\,<$\,6.0~PeV and a 90\% CL upper limit of about 1467
times the predicted fluence as shown in figure~\ref{fig:limits} (left) was set.

\section{Conclusions}
\label{sec:conclusions}
A search for \mneus\ in correlation with the prompt emission of
\gammaraybursts\ using the data taken with the \antares\ detector
during the first year of operation has been presented.
During the period considered, the detector was less than half its
final size. 
The \ngrbs\ GRBs that were examined for neutrino emission were
selected from the GRB observations by satellite instruments.
No correlations between neutrinos and the selected GRBs have been
found. 
Upper limits have been obtained on the fluence of \mneus\ from the  
\ngrbs\ GRBs and on the diffuse \mneu\ flux for different models.

A neutrino telescope in the Mediterranean Sea is well suited to detect
high-energy neutrinos in correlation with GRBs spread over a wide
region of the sky, including in particular the Southern hemisphere.
The low background is the result of the short duration of GRBs and
the excellent angular resolution.
The \antares\ detector was completed mid 2008 and is 2.5 times bigger
than the detector configuration considered in this analysis. 
Since the completion of the detector, on average 250 GRBs 
per year have been detected in the Southern hemisphere and at least five 
more years of data taking are foreseen. 
With this, a large sample of GRBs is available for further analysis, 
complementary to the IceCube field of view and energies. 
The next-generation neutrino telescope KM3NeT~\cite{km3net},
to be built in the Mediterranean Sea, will surpass the
\antares\ sensitivity by two orders of magnitude.
Once operational it will probe GRB models with unprecedented
sensitivity, and will push the boundaries towards new discoveries. 

\acknowledgments
The authors acknowledge the financial support of the funding agencies:
Centre National de la Recherche Scientifique (CNRS), Commissariat
\'{a} l'\'{e}nergie atomique et aux \'{e}nergies alternatives  (CEA), Agence
National de la Recherche (ANR), Commission Europ\'{e}enne (FEDER fund 
and Marie Curie Program), R\'{e}gion Alsace (contrat CPER), R\'{e}gion 
Provence-Alpes-C\^{o}te d'Azur, D\'{e}\-par\-tement du Var and Ville de 
La Seyne-sur-Mer, France; Bundesministerium f\"{u}r Bildung und Forschung 
(BMBF), Germany; Istituto Nazionale di Fisica Nucleare (INFN), Italy; 
Stichting voor Fundamenteel Onderzoek der Materie (FOM), Nederlandse 
organisatie voor Wetenschappelijk Onderzoek (NWO), the Netherlands; 
Council of the President of the Russian Federation for young scientists 
and leading scientific schools supporting grants, Russia; National 
Authority for Scientific Research (ANCS - UEFISCDI), Romania; Ministerio 
de Ciencia e Innovaci\'{o}n (MICINN), Prometeo of Generalitat Valenciana 
and MultiDark, Spain; Agence de l'Oriental, Morocco. We also
acknowledge the technical support of Ifremer, AIM and Foselev Marine
for the sea operation and the CC-IN2P3 for the computing facilities.

\end{document}